\documentstyle[11pt,newpasp,twoside,epsf,psfig]{article}
\def\masyr{\,{\rm mas}\,{\rm yr}^{-1}}
\def\kpc{\,{\rm kpc}}
\def\Gyr{\,{\rm Gyr}}
\def\fracj#1#2{{\textstyle{#1\over#2}}}
\def\kms{\,{\rm km}\,{\rm s^{-1}}}\def\msun{\,{\rm M}_\odot}
\begin{document}
\title{Dynamics of the Galaxy's Satellites}
 \author{James Binney}
\affil{Oxford University, Theoretical Physics, Keble Road, Oxford, OX1 3NP,
U.K.}

\begin{abstract}
The Milky Way's satellites provide unique information about the density of
the Galactic halo at large radii. The inclusion of even a few rather
inaccurate proper motions resolves an ambiguity in older mass estimates in
favour of higher values. Many of the satellites are concentrated into
streams. The dynamics of the Magellanic Stream provided an early indication
that the halo reaches out to beyond $100\kpc$. Tidal forces between the
Clouds are currently disturbing the Clouds' internal dynamics.  One would
expect this damage to worsen rapidly as the tidal field of the MW excites
the eccentricity of the Clouds' mutual orbit. This process, which has yet to
be completely modelled, is important for understanding the degree of
self-lensing in searches for gravitational lensing events. The Sagittarius
Dwarf galaxy very likely contributes significantly to the Galactic warp. The
direction of the warp's line of nodes is incorrectly predicted by the
simplest models of the Dwarf's orbit. More sophisticated models, in which a
complex distribution of stripped dark matter is predicted, may be more successful.
\end{abstract}

\section{Introduction}

The flat rotation curves of many external galaxies have convinced us that
the space around galaxies is filled with dark matter of some sort.
Elucidating the nature of this matter is one of the central problems of
contemporary astronomy. By learning more about the dynamics of the
$\sim100\kpc$ around galaxies, we may determine what this matter is.

Direct measurement of the Milky Way's circular-speed curve becomes
problematic beyond $R_0$ (e.g., Binney \& Dehnen, 1997). We can, however,
probe the dynamics of the outer Milky Way with observations of objects that
are too faint to be studied in much detail around external galaxies, so in
some respects we know more about the Galactic dark halo than about the dark
halo of any other galaxy.

\section{Equilibrium Spherical Models}

\begin{figure}
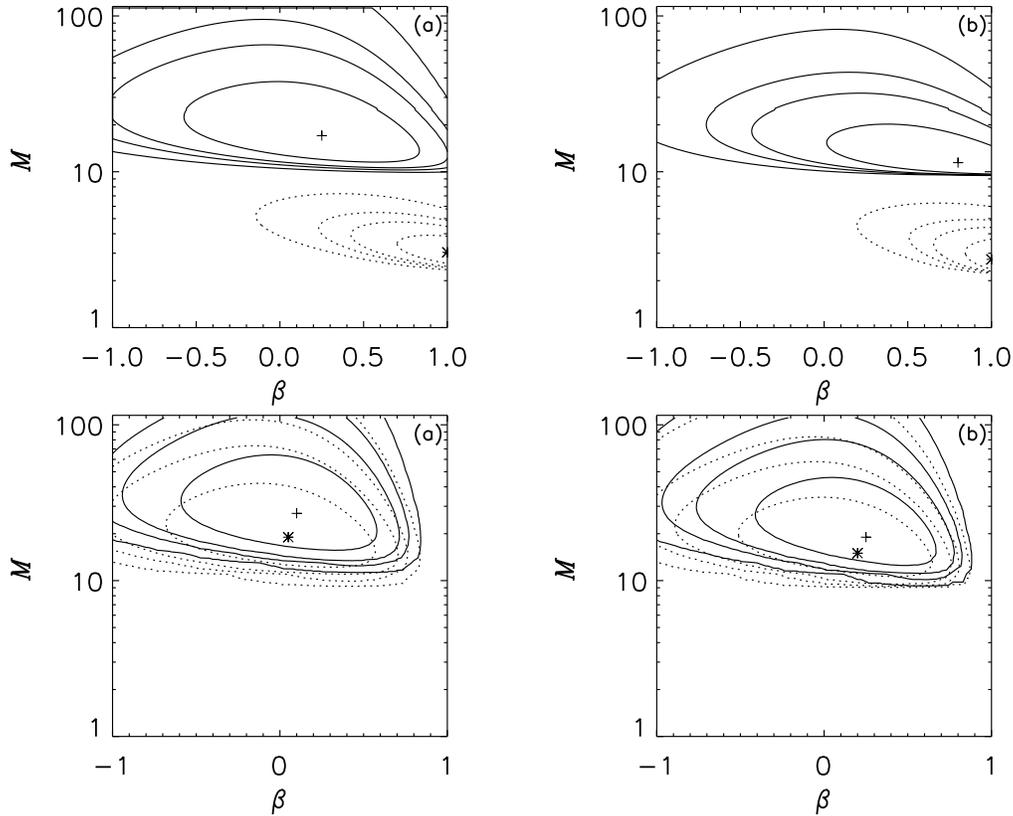

\centerline{\psfig{file=wilk37.ps,width=\hsize}}
\vskip2pt
\centerline{\psfig{file=wilk40.ps,width=\hsize}}
\caption{Contours of equal likelihood for mass models of the Galactic halo,
with (full) and without (dotted) data for Leo I. When five proper motions
are used, one obtains the lower  panels, while the upper panels are obtained
when only line-of-sight velocities are used. The left and right panels
derive from models with different assumptions about the halo's radial
density profile. The parameter $\beta$ controls the anisotropy of the
model's velocity distribution, being positive for radial anisotropy. [From
Wilkinson \& Evans (1999).]\label{wilkfig}}

\end{figure}

In the simplest picture, the dark halo is spherical and phase mixed. Over
the last decade and a half many attempts have been made to constrain the
mass of such a halo from observations of the line-of-sight velocities of
distant globular clusters and dwarf satellite galaxies (Little \& Tremaine
1987; Zaritsky et al.\ 1989; Kulessa \& Lynden-Bell 1992; Kochanek 1996).
The masses inferred in these studies depend strongly on whether or not the
most distant satellite, Leo I, is assumed to be bound to the Milky Way.
Wilkinson \& Evans (1999) have revisited this problem and shown that when
radial velocities are complemented by proper motions, consistent mass
estimates are obtained with or without Leo I. Figure \ref{wilkfig}
illustrates this result.  In the top panels the likelihood peaks at models
that differ in mass by a factor of 2 depending on whether Leo is included
(full contours) or not. This is the case in which only line-of-sight
velocities are employed. In the lower panels, which include five proper
motions, the likelihood peaks at masses that differ insignificantly 
whether Leo I is included or not.  Quantitatively, the mass required inside
$50\kpc$ is $M_{50}=(5.45\pm0.15)\times10^{11}\msun$, while that required
inside $100\kpc$ is $M_{100}=(10\pm0.4)\times10^{11}\msun$. Large though
they are, these masses are still less than half the mass of the Local Group,
$M_{\rm LG}=(6\pm2)\times10^{12}\msun$ (Schmoldt \& Saha 1998), so most of
the mass of the Universe lies outside the halos of galaxies, just as several
cosmological arguments predict.

It is to be expected that proper motions provide considerable leverage on
the problem because the Sun lies close to the centre of the halo, with the
consequence that the tangential motions of distant objects are virtually
unconstrained by line-of-sight velocities. The upper panels in Figure
\ref{wilkfig} show that in the absence of proper motions, highly radially
anisotropic models are strongly favoured, and with Leo I excluded these
models have rather low masses. When proper motions are included, these
anisotropic models become less likely, and larger masses are required. What
is perhaps remarkable is that such dramatic shifts in likelihood are
obtained with the extant proper motions, in which the errors are large -- of
order $0.3\masyr$ in proper motions that lie in the range $2.7$ to
$0.3\masyr$. Over the coming decade, as results from the upcoming generation
of astrometric satellites (DIVA, FAME, SIM and GAIA) become available, the
errors in these proper motions will be dramatically reduced, and many
additional proper motions will become available. The data will then
determine the mass profile of the Galactic halo to high precision.

\section{Streams and Infall}

At $100\kpc$ from the Galactic centre, the dynamical time is $t_{\rm
dyn}=\fracj12\pi\sqrt{R^3/GM}\sim0.8\Gyr$, so we cannot expect the outer halo
to be dynamically relaxed. In particular, at large radii we should expect
material to be falling in to the Milky Way for the first time (Gunn \& Gott,
1972). Recently, Blitz et al.\ (1999) and Braun \& Burton (1999, 2000)
have powerfully restated the case that compact
high-velocity clouds are such infalling material. Indirect evidence for past infall comes from the spread
($\sim7\Gyr$) in the ages of globular clusters (Stetson, vandenBergh \&
Bolte, 1996), the classic G-dwarf
problem (van den Bergh, 1962; Binney \& Merrifield 1998, \S10.7.2), the counter-rotation
of the system that is formed by globular clusters with metallicities
$-1.7<\hbox{[Fe/H]}<-1.3$ (Rodgers \& Paltoglou 1984), and the existence of
satellite streams.

Lynden-Bell (1976) pointed out that the Galaxy's satellites appear to lie
along a few great circles in the sky. Lynden-Bell \& Lynden-Bell (1995)
markedly refined this conjecture by adding kinematic data to the analysis.
They identified four high-quality streams with the following members:

\begin{itemize}

\item Fornax, Pal 14, Pal 15, (Eridanus ?)

\item Magellanic Clouds, Draco, Ursa Minor, (Sculptor, Carina ?)

\item NGC 2419, NGC 7006, Rup 106, (Ter 7, Sagittarius Dwarf ?)

\item Pal 2, Arp 2, Sagittarius Dwarf, (Ter 7 ?)

\end{itemize}

Interest in these streams is currently very high as it has been suggested
(Johnston et al., 1999) that the Galactic potential could be accurately
determined if the space velocities of a few objects in each stream could be
reliably measured. SIM and GAIA will make the measurements, but I am
optimistic that the potential will have been precisely mapped before SIM or
GAIA data become available through detailed modelling of observations of
more numerous classes of halo tracers, such as blue horizontal-branch stars,
for which less complete data are available for any individual object (e.g.,
Dehnen \& Binney, 1996).

\subsection{The Magellanic Stream}

Far and away the best studied stream is that of the Magellanic Clouds, which
includes a stream of HI that reaches nearly half way across the southern
sky. Murai \& Fujimoto (1980) first gave the currently accepted model of the
Magellanic Stream.  From the fact that the LMC and SMC form a binary system,
they deduced that the Clouds are near pericentre rather than near apocentre, as
had been previously thought. It followed that the gaseous stream was
trailing the Clouds, and an extensive Galactic halo was required to generate
the observed line-of-sight velocities along the stream. Any doubt as to the
essential correctness of the Murai \& Fujimoto model has been eliminated by
the measurement of the LMC's proper motion (Jones, Klemola \& Lin 1994;
Kroupa \& Bastian 1997), which  the model correctly predicted.

Most modelling of the Magellanic Stream has used test particles, with the
effect of dynamical friction against the Galactic halo (which is dynamically
important) added analytically. Gardiner \& Noguchi (1996) have usefully
extended this kind of analysis by representing the SMC by a self-consistent
$N$-body model. Their simulations give a convincing picture of the formation
of the Magellanic bridge and stream from material that has been tidally torn
from the disk of the SMC. This kind of modelling is important because we
need more convincing estimates of the amount of gravitational self-lensing
to be expected within the Clouds (Sahu, 1994; Kerins \& Evans, 2000), and
this depends strongly on the vertical structure of the Clouds, which is in
turn going to depend on the tidal distortion of one Cloud by the other. The
model of Gardiner \& Noguchi (1996) does not predict as much self-lensing
(Graff \& Gardiner 1999) as now seems probable (Kerins \& Evans, 2000), but
the parameter space to be explored, which includes the orientation of the
disk of each cloud and the mutual orbit of the Clouds relative to the orbit
of the Clouds' barycentre around the Milky Way, is large and has yet to be
fully explored.

\subsection{Dynamics of the Sagittarius Dwarf}

Since its discovery $16\kpc$ behind the Galactic centre (Ibata et al.\
1994), the Sagittarius dwarf galaxy has attracted a good deal of attention.
Like the Clouds, the Dwarf is on a nearly polar orbit, but the plane of its
orbit is nearly orthogonal to that of the Clouds' orbit. The period of the
Dwarf's orbit is remarkably short, $\la1\Gyr$, and there is general
agreement that it is currently being torn apart by the Galaxy's tidal force.
How long has the Dwarf been on this exposed orbit?

Several studies have concluded that the Dwarf could not survive for a Hubble
time on its current orbit. The exceptions to this rule are Ibata \& Lewis
(1998) and Helmi \& White (2000). Ibata \& Lewis showed that a very nearly
homogeneous dark-matter halo for the Dwarf of mass $1.2\times10^9\msun$ can
survive for a Hubble time, although it has by then been tidally stripped of
$\sim60\%$ of its mass.  Helmi \& White show that models with initial mass
$0.57\times10^9\msun$ and $1.2\times10^9\msun$ yield reasonable matches to
the observations after a Hubble time, although they have by then been
stripped of almost 90\% of their mass. Whether the stripped mass is
associated with light and is optically traceable depends on the assumed
radial variation of the mass-to-light ratio within the initial model. 

Could the Dwarf have formed on  its present orbit, along which tidal
shredding is such an efficient process? Or has the Dwarf migrated to
its current short-period orbit from a longer-period and safer one? Zhao
(1998) made the ingenious proposal that the Dwarf was deflected onto its
current orbit by the Magellanic Clouds as the systems passed one another
near a Galactic pole. The problem with this idea is to understand how the
Dwarf could have been deflected by a large angle on encountering an object
with a similarly soft potential at a speed of $\sim300\kms$.

Another mechanism by which the Dwarf could have moved to a short-period
orbit is dynamical friction. Its current mass is so small that it will now be
suffering negligible frictional drag. But we have seen that it must
currently be losing mass rapidly, and will have been more massive in the
past. Jiang \& Binney (2000) show that there is a one-parameter family of
initial configurations of the Dwarf that evolve into something like its
present configuration over a Hubble time. At one extreme we have a Dwarf of
mass $\sim10^{11}\msun$ starting from Galactocentric radius $\sim250\kpc$.
At the other extreme the initial mass is $\sim1.2\times10^9\msun$ and the
Dwarf starts from $\sim60\kpc$, very much as in the models of Ibata \& Lewis
and Helmi \& White.

\section{Warp of the Milky Way}

On any model in which the Dwarf has an effectively polar orbit, virtually
all the Dwarf's initial mass is now in a polar annulus.  The mass of this
annulus can be as little as $10^9\msun$ or as much as $\sim10^{11}\msun$,
and its angular momentum is uncertain by even more than two orders of
magnitude, because the higher-mass rings will have larger mean radii. An
angular-momentum detector would seem to be the best way of distinguishing
between these possibilities. The Galactic disk is just such a detector in
that it becomes warped in response to the addition of off-axis angular
momentum (Jiang \& Binney, 1999). The way the ring of matter stripped from
the Dwarf affects the Galactic disc depends sensitively on how nearly polar
the Dwarf's orbit is: if it were exactly polar, there would be no torque acting
between the ring and the Galaxy, and no distortion of the disk. The
inclination of the Dwarf's orbit is not accurately known, but the pole of
its orbit is thought to lie near $(l,b)=(85\deg,25\deg)$. With this orientation
the ring is pulling the northern disk down and the southern disk up, and the
vector of the torque on the disk points from the Galactic centre towards the
Sun. Because it rotates clockwise, the disk's angular-momentum vector points
to the SGP, and the Dwarf should be rotating it up towards the Sun.
Consequently, the line of nodes should lie in the direction $l=90\deg$,
rather than along $l=0\deg$ as observed. 

This failure of the model to generate the correct orientation for the warp's
line of nodes is frustrating because the magnitude of the predicted effect
is about right. To give a concrete example, consider the effect upon the
disc of a ring of mass $2\times10^{9}\msun$ and radius $25\kpc$ whose polar
axis lies at $(l,b)=(90\deg,25\deg)$. If we adopt the simplest model of the
Galactic potential, $\Phi_{\rm MW}=v_c^2\ln r$ with $v_c=220\kms$, we find
that the angular momentum vector of a star on a circular orbit of radius
$16\kpc$ shifts towards the current solar position at a rate 
$0.43\deg\Gyr^{-1}$, so after $10\Gyr$ the star would oscillate $\sim1.2\kpc$
around its original orbital plane. The HI data require just such excursions
(e.g., Fig.~9.24 of Binney \& Merrifield, 1998).

Two points should be considered when trying to resolve the problem posed by
the observed line of nodes. The first is the Magellanic Clouds: their orbit
is almost perpendicular to that of the Dwarf, so do they not generate a warp
whose line of nodes agrees nicely with that observed? The answer is `no' for
two reasons. First, the orbit of the Clouds is so nearly polar [its pole
lies near $(l,b)=(190\deg,3\deg)$] that a uniform ring with this orientation
would apply only a very small torque to the disk. The Clouds are never the
less capable of inducing a warp because they have not yet been uniformly
smeared around a ring, and the period of their orbit is long enough that, in
a useful approximation, they can be considered stationary. At their current
location they are pulling the southern disc down and thus applying a torque
that acts on the same line as the torque from the Dwarf's ring, but has
opposite sign (Garcia-Ruiz, Kuijken \& Dubinski, 2000). Consequently, the
associated line of nodes is parallel to that predicted by the Dwarf.

Unless the mass of the Dwarf is at the bottom end of expectations, the Dwarf
must contribute non-negligibly to the Galactic warp.  The problem with the
line of nodes just described may be resolved by precession of the Dwarf's
stripped material. Since it is torquing the disk, this material must itself
be precessing, and the current position of the Dwarf's luminous core may
not be a good guide to the current location of the mass of dark matter that
was stripped some Gyr ago. At present  no model is sufficiently
sophisticated to enable one to evaluate this idea quantitatively.

\section{Conclusion}

The Milky Way's satellites provide a wealth of information about the extent
and history of our very typical $L^*$ galaxy. Natural extensions of existing
models of the dynamics of streams promise a rich harvest of insights into
some of the most important questions in astronomy.

\end{document}